\documentclass[12pt]{article}

\def\la{\lambda}

\def\S{{\hat S}}
\def\V{{\hat V}}
\def\T{{\hat T}}
\def\a{{\alpha}}
\def\b{{\beta}}

\def\be{\begin{equation}}
\def\ee{\end{equation}}

\usepackage{amsmath}
\usepackage{latexsym}

\newtheorem{theorem}{Theorem}

\begin{document}

\title{On factorization and solution of multidimensional linear partial 
differential equations}

\author{S.P.~Tsarev\thanks{On leave from: Krasnoyarsk State Pedagogical
  University, Russia. Partial financial support was provided by
   RFBR grants 04-01-00130 and 06-01-00814.}\\[1ex]
   Department of Mathematics \\
 Technische Universit\"at Berlin \\
Berlin, Germany \\[1ex] 
e-mails: \\
\texttt{tsarev@math.tu-berlin.de} \\ 
\texttt{sptsarev@mail.ru}\\
}
   
\date{September 2006}   
 
   \maketitle

\medskip

\begin{abstract}
    We describe a method of obtaining closed-form complete solutions of 
  certain second-order linear partial  differential equations with more than two independent variables. This method generalizes
the classical method of Laplace transformations of
second-order hyperbolic equations in the plane and is based on an idea given by 
Ulisse Dini in 1902.
\end{abstract}

\section{Introduction}

Factorization of linear partial differential operators
\mbox{(LPDOs)} is often used in modern algorithms for solution of
the corresponding differential equations. In the last 10 years
a number of new  modifications and generalizations of classical algorithms for
factorization of LPDOs were
given (see e.g.\ \cite{Athorne,grig1,grig2,L-Sch-Ts,minwu,SW,ts98,ts05}).
 Such results have also close links with the theory of explicitly integrable nonlinear partial differential equations, cf.\ \cite{anderson,sokolov,S-Zh}.

As one can see from simple examples (cf.\ \cite{ts98} and Section~\ref{lap} below) a ``naive''  definition of factorization of a given LPDO $\hat L$ as its representation
as a composition  $\hat L=\hat L_1\circ \hat L_2$ of lower-order operators does not enjoy good properties and in general is not  related to existence of a complete closed-form solution.

On the other hand, for  second-order hyperbolic linear equations in the plane  we have a well established and deep theory of ``generalized'' factorization. This theory is known since the end of the XVIII century under the name of
Laplace cascade method or Laplace transformations.  As proved in  \cite{ts98}, existence of a complete solution of a given second-order hyperbolic equation in the plane in explicit form is equivalent to some ``generalized factorizability'' of the corresponding operator which in turn is equivalent to finiteness of the chain of Laplace transformations ending in a ``naively'' factorizable operator. We give a short account of this method in Section~\ref{lap}.

There were some attempts to generalize Laplace transformations for higher-order operators or larger number of independent variables, both in the classical time \cite{LeRoux,pisati,Petren} and in the last decade
\cite{Athorne,ts05}.
A general  definition of generalized factorization comprising all known practical methods was given in \cite{ts98}. Unfortunately the theoretical considerations of  \cite{ts98} did not provide any algorithmic way of establishing generalized factorizability of a given LPDO.

In this paper we move a bit further, extending the algorithmic methods  for generalized factorization to the case of second-order operators in the space of more than two independent variables following an approach proposed by Ulisse Dini in 1902 \cite{dini1,dini2}.

The paper is organized as follows. In the next Section we give an exposition of the classical theory of Laplace transformations.
In Section~\ref{diniex} we work out an example which demonstrates the idea of Dini transformations and in Section~\ref{dinigen} we prove a new general result showing that this idea gives a practical method applicable to arbitrary hyperbolic second-order linear equation in the three-dimensional space provided the principal symbol of the operator factors.
The last Section is devoted to discussion of algorithmic problems encountered in the theory of Laplace and Dini transformations, their relations to the given in \cite{ts98} theoretical basis. Some conjectures on possibility of generalized factorization and existence of complete solutions in closed form are given.

\section{Laplace and generalized Laplace transformations}\label{lap}

The cascade method of Laplace (also called the method of Laplace transformations)
is until the present date the most general method of finding closed-form complete solutions of hyperbolic second-order linear partial  differential equations with two independent variables. 
Here we briefly sketch this classical theory in a form convenient for our purpose. The
complete account may be found for example in
\cite{forsyth,gour-l}.

Let a general second-order linear operator $L$
with two independent variables $x$ and $y$ be given:
\begin{equation}\label{L2}
  \hat L= \sum_{i=0}^2p_{i}\hat D_x^i\hat D_y^{2-i} + a_1(x,y)\hat D_x +
  a_2(x,y)\hat D_y + c(x,y),
\end{equation}
$p_i=p_{i}(x,y)$, $\hat D_x = \partial / \partial x$, $\hat D_y = \partial / \partial y$.
Hereafter we will always suppose that the operator (\ref{L2})
is strictly hyperbolic, i.e.\ the characteristic equation
$\la^2p_0-\la p_1+p_2=0$ for the principal symbol of
(\ref{L2}) has two distinct real roots $\la_1(x,y)$, $\la_2(x,y)$,
so we can introduce two first-order characteristic operators 
$\hat X_i = m_i(x,y)\hat D_x +
n_i(x,y)\hat D_y$, $i=1,2$, $m_i/n_i=\la_i$
($\hat X_i$ are defined up to rescaling $\hat X_i \rightarrow
\gamma_i(x,y)\hat X_i$).

The corresponding equation $Lu=0$ now may be rewritten in 
one of two {\em characteristic forms}: 
\begin{equation}\label{4}
  \begin{array}{l}
  (\hat X_1\hat X_2 + \alpha_1\hat X_1 + \alpha_2\hat X_2 +
  \alpha_3)u
  = \\
(\hat X_2\hat X_1 + \overline\alpha_1\hat X_1 + \overline\alpha_2\hat X_2 +
\alpha_3)u=0,
\end{array}
\end{equation}
where $\alpha_i=\alpha_i(x,y)$. Since the operators $\hat X_i$ do not necessarily
commute we have to
take into consideration in (\ref{4}) and everywhere below the {\em
commutation law}
\begin{equation}\label{cl}
  [\hat X_1,\hat X_2] = \hat X_1\hat X_2- \hat X_2\hat X_1 = P(x,y)\hat X_1 +Q(x,y)\hat X_2.
\end{equation}
Using the {\em Laplace invariants} of the operator (\ref{4})
$$h=\hat X_1(\alpha_1) +\alpha_1\alpha_2 -\alpha_3, \quad 
k=\hat X_2(\overline\alpha_2) +\overline\alpha_1\overline\alpha_2-
\alpha_3,$$ we represent the original operator $\hat L$ in two possible
partially factorized forms:
\begin{equation}\label{pf}
 \hat L = (\hat X_1+\alpha_2)  (\hat X_2+\alpha_1)-h=
  (\hat X_2+\overline\alpha_1) (\hat X_1+\overline\alpha_2) -k.
\end{equation}
From these forms we see that the equation $\hat L u =0$ is equivalent
to any of the following two first-order systems
\begin{equation}\label{5}
\!\!(S_1)\!:\!\left\lbrace  \begin{array}{l} \!\!\hat X_2u=-\alpha_1u
+v,
\\ \!\! \hat X_1v =hu-\alpha_2v.
\end{array}\right. \!\!\!\Leftrightarrow
\!(S_2):\!\left\lbrace  \begin{array}{l}
 \hat X_1u=-\overline\alpha_2u +w, \\
  \hat X_2w =ku-\overline\alpha_1w.
\end{array}\right.\!\!\!\!
\end{equation}
If at least one of the Laplace invariants $h$ or
$k$ vanishes identically, then the operator $\hat L$ factors (in the ``naive''
way) into composition of two first-order operators and the corresponding system in
(\ref{5}) becomes triangular. So the problem of integration of the original second-order equation is reduced to a much easier problem of integration of linear first-order equations. The latter problem is essentially reducible to finding the complete solution of a (nonlinear!) ODE, see below Section~\ref{concl}.

 If $h \neq 0$, $k \neq 0$, one can take  one of the systems (\ref{5}) (to fix the notations we choose
the left system $(S_1)$), express $u$ using the second equation of
$(S_1)$
\begin{equation}\label{difsub}
u = (\hat X_1 v +\alpha_2 v)/h
\end{equation}
and substitute this expression into the first equation of  $(S_1)$ in (\ref{5}); as the result one obtains a {\em $X_1$-transformed}
equation $\hat L_{(1)} v =0$. It has {\em different} Laplace invariants
 (cf.\ \cite{anderson})
$$   
\begin{array}{l}
 h_{(1)} =\hat X_1(2\alpha_1 - P)  -  \hat X_2(\alpha_2)  - \hat X_1\hat X_2\ln h   +Q\hat X_2\ln h    - \alpha_3 +(\alpha_1-P)(\alpha_2-Q)
 \\
  =2h-k \hat X_1\hat X_2\ln h + Q\hat X_2\ln h +\hat X_2(Q) -\hat X_1(P)
+2PQ,
 \\[0.5em]
k_{(1)}=h.
\end{array}
$$  
If $h_{(1)}=0$, we can solve this new equation in quadratures and
using the same differential substitution (\ref{difsub}) we obtain
the complete solution of the original equation $\hat L u=0$.

 If again $h_{(1)} \neq 0$, apply this
$X_1$-transformation several times, obtaining a sequence of
second-order operators $\hat L_{(2)}$, $\hat L_{(3)}$, \ldots\ of
the form (\ref{4}). If on any step we get $h_{(k)}=0$, we solve
the corresponding equation $\hat L_{(k)} u_{(k)} =0$ in quadratures
and, using the differential substitutions (\ref{difsub}), obtain
the complete solution of the original equation. Alternatively one
may perform {\em $\hat X_2$-transformations}: rewrite the original
equation in the form of the right system $(S_2)$ in (\ref{5}) and
using the substitution $u = (\hat X_2 w +\overline\alpha_1 w)/k$ obtain the equation $\hat L_{(-1)} w =0$ with Laplace invariants
\begin{equation}\label{yl}
\begin{array}{l}
h_{(-1)}=k, \\[0.5em]
 k_{(-1)}= 2k-h - \hat X_2\hat X_1\ln k - P\hat X_1\ln k
 +\hat X_2(Q)  -\hat X_1(P) +2PQ.
\end{array}
\end{equation}
In fact this $\hat X_2$-transformation is a reverse of the
$\hat X_1$-trans\-for\-ma\-tion up to a gauge transformation (see
\cite{anderson}). So we have (infinite in general) chain of
second-order operators
\begin{equation}\label{ch}
   \ldots \stackrel{\hat X_2  }{\leftarrow} \hat L_{(-2)} \stackrel{\hat X_2  }{\leftarrow}
   \hat L_{(-1)}\stackrel{\hat X_2  }{\leftarrow}   \hat L \stackrel{\hat X_1}{\rightarrow}
    \hat L_{(1)} \stackrel{\hat X_1}{\rightarrow} \hat L_{(2)} \stackrel{\hat X_1}{\rightarrow}
     \ldots
\end{equation}
 As one may prove (see e.g.\ \cite{gour-l}) if the chain
(\ref{ch}) is finite in both directions (i.e.\ we have
$h_{(N)}=0$, $h_{(-K)}=0$ for some $N\geq 0$, $K\geq 0$) one may
obtain a quadrature-free expression of the general solution of the
original equation:
\begin{equation}\label{XY}
   u =   c_0F + c_1F' +  \ldots  +   c_NF^{(N)}   +
d_0 G + d_1 G' +
   \ldots +  d_{K-1} G^{(K-1)}
\end{equation}
with definite $c_i(\overline x,\overline y)$, $d_i(\overline
x,\overline y)$ and  $F(\overline x)$, $G(\overline y)$
--- two arbitrary functions of the characteristic variables and
vice versa: existence of ({\em a priori} not complete) solution of
the form (\ref{XY}) with arbitrary functions $F$, $G$ of
characteristic variables implies $h_{(s)}=0$, $h_{(-r)}=0$ for
some $s \leq N$, $r \leq K-1$. So {\em minimal differential
complexity} of the answer (\ref{XY}) (number of terms in it) is
equal to the number of steps necessary to obtain vanishing Laplace
invariants in the chain (\ref{ch}) and consequently
naively-factorable operators. 
 If (\ref{ch}) is finite in one direction only, one can still obtain a closed-form
expression for the complete solution of the original equation; however,
it will have {\em one} of the free functions $F$ or $G$ inside  a quadrature expession.
More details and complete proofs of these statements
may be found in \cite{forsyth,gour-l} for
the case $\hat X_1=\hat D_{x}$, $\hat X_2=\hat D_{y}$, for the general case cf.\
\cite[p. 30]{gour-l} and \cite{anderson}.

{\it Example 1.} As a straightforward computation shows, for the
equation $u_{xy} - \frac{n(n+1)}{(x+y)^2}u=0$  the chain
(\ref{ch}) is symmetric ($h_{(i)} = h_{(-i-1)}$) and has length
$n$ in either direction. So the complexity of the answer
(\ref{XY}) may be very high and depends on some arithmetic
properties of the coefficients of the operator $\hat L$; for the
equation $u_{xy} - \frac{c}{(x+y)^2}u=0$ the chain (\ref{ch})
 will be infinite unless the constant $c=n(n+1)$.

Recently a generalization of this classical method was given
in \cite{ts05}. In is applicable to strictly hyperbolic linear equations of arbitrary order with two independent variables $x$, $y$ only.

In \cite{dini1} a simple generalization of Laplace transformations formally applicable to some second-order operators in the space of arbitrary dimension was proposed. Namely, suppose that such an operator
$\hat L$ has its principal symbol 
$$Sym=\sum_{i_1,i_2}a_{i_1i_2}(\vec x)\hat D_{x_{i_1}}\hat D_{x_{i_2}}$$
which factors (as a formal polynomial in formal commutative variables
$\hat D_{x_{i}}$) into product of two first-order factors: $Sym=\hat X_1\hat X_2$ (now $\hat X_j=\sum_{i}b_{ij}(\vec x)\hat D_{x_{i}}$) and moreover the complete operator $\hat L$ may be written at least in one of the forms given in (\ref{4}). This is very restrictive since the two tangent vectors corresponding to the first-order operators $\hat X_i$
no longer span the complete tangent space at a generic point $(\vec x_0)$. (\ref{cl}) is also possible only in the case when these two vectors give an \emph{integrable} two-dimensional distribution of the tangent subplanes in the sense of Frobenius, i.e.\ when one can make a change of the independent variables $(\vec x)$ such that $\hat X_i$ become parallel to the coordinate plane $(x_1,x_2)$; thus in fact we have an operator $\hat L$ with only $\hat D_{x_{1}}$,  $\hat D_{x_{2}}$ in it and we have got no really significant generalization of the Laplace method.
If one has only (\ref{4}) but (\ref{cl}) does not hold one can not perform more that one step in the Laplace chain (\ref{ch}) and there is no possibility to get an operator with a zero Laplace invariant (so naively factorizable and solvable).

In the next section we demonstrate, following an approach proposed by U.~Dini in another paper \cite{dini2}, that one can find a better analogue of Laplace transformations for the case when the dimension of the underlying space of independent variables is greater than two. Another particular special transformation was also proposed in \cite{Athorne}, \cite{YA}; it is applicable to systems whose order coincides with the number of independent variables. The results of
\cite{Athorne}, \cite{YA} lie beyond the scope of this paper.

\section{Dini transformation: an example}\label{diniex}

Let us take the following equation: 
\begin{equation}\label{dex}
 Lu = (\hat D_x\hat D_y + x \hat D_x\hat D_z - \hat D_z)u =0.
\end{equation}
It has three independent derivatives $\hat D_x$, $\hat D_y$, $\hat D_z$,
so the Laplace method is \emph{not} applicable.
On the other hand its principal symbol splits into product of two
first-order factors: $\xi_1\xi_2 + x \xi_1\xi_3 =\xi_1(\xi_2+x\xi_3)$.
This is no longer a typical case for hyperbolic operators 
in dimension~$3$; we will use this special feature introducing
two characteristic operators
$\hat X_1=\hat D_x$, $\hat X_2=\hat D_y + x \hat D_z$. We have again a nontrivial
commutator  $[\hat X_1,\hat X_2] =  \hat D_z= \hat X_3$. The three operators $\hat X_i$ span the complete tangent space in every point $(x,y,z)$.
Using them one can represent the original second-order operator in one of two
partially factorized forms:
$$ L = \hat X_2\hat X_1 - \hat X_3 =  \hat X_1\hat X_2 - 2\hat X_3.$$
Let us use the first one and transform the equation into a system
of two first-order equations:
\begin{equation}\label{Dini2e}
 Lu=0 \Longleftrightarrow 
   \left\lbrace  \begin{array}{l}
\hat X_1 u = v, \\
\hat X_3 u = \hat X_2 v.
\end{array}\right.
\end{equation}
Here comes the difference with the classical case $dim=2$: we can not
express $u$ as we did in (\ref{difsub}). But we have another obvious
possibility instead: cross-differentiating the left hand sides of 
(\ref{Dini2e}) and using the obvious identity
$[\hat X_1,\hat X_3] = [ \hat D_x, \hat D_z]=0$ we get
$  \hat X_1 \hat X_2v =   \hat D_x (\hat D_y + x\hat D_z) v = \hat X_3 v=\hat D_z v $ or
$ 0=\hat D_x (\hat D_y + x\hat D_z) v - \hat D_z v = (\hat D_x \hat D_y + x \hat D_x \hat D_z) v
 = (\hat D_y + x\hat D_z)  \hat D_x v = \hat X_2\hat X_1 v$.
 
This is precisely the procedure proposed by Dini in \cite{dini2}.
Since it results now in another second-order equation which is 
``naively'' factorizable we easily find its complete solution:
$$v= \int \phi(x,xy-z) \, dx + \psi(y,z)$$ 
where $\phi$ and $\psi$ are two arbitrary functions of
two variables each; they give the general solutions
of the equations $\hat X_2\phi=0$, $\hat X_1\psi=0$.

Now we can find  $u$:
$$ u= \int \Big(v\, dx  + (\hat D_y + x\hat D_z)v\, dz \Big)+ \theta(y),
$$
where an extra free function $\theta$ of one variable appears
as a result of integration in (\ref{Dini2e}).

So we have seen that such {\em Dini transformations}
(\ref{Dini2e}) in some cases may produce a complete solution in explicit form for a non-trivial three-dimensional equation (\ref{dex}).
This explicit solution can be used to solve initial value problems
for (\ref{dex}).

\section{Dini transformation: a general result for $dim=3$, $ord=2$}\label{dinigen}

Dini did not give any general statement on the range of applicability of 
his trick. In this section we investigate this question. Obviously one can make different transformations similar to that demonstrated in the previous section, here we concentrate on the simplest case of second-order linear equations with three independent variables
{\em whose principal symbol factors}.
\begin{theorem}
 Let an operator $L=\sum_{i+j+k\leq 2}a_{ijk}(x,y,z)\hat D_x^i\hat D_y^j\hat D_z^k$ has
a factorizable principal symbol: $\sum_{i+j+k= 2}a_{ijk}(x,y,z)\hat D_x^i\hat D_y^j\hat D_z^k= \S_1\S_2$ (modulo lower-order operators) with  (non-commuting) first-order operators $\S_1$, $\S_2$; $\S_1 \neq \la(x,y,z)\S_2$. 
Then in the generic case there exist two Dini transformations $L_{(1)}$, $L_{(-1)}$ of $L$.
\end{theorem}
\textsl{Proof.}
One can represent $L$ in two possible ways:
\begin{equation}\label{th-1}
L=\S_1\S_2 + \T + a(x,y,z) =\S_2\S_1 + \hat U + a(x,y,z)
\end{equation}
with some first-order operators $\T$, $\hat U$. 
We will consider the first one
obtaining a transformation of $L$ into an operator $L_{(1)}$ of similar form. 

In the  generic case the operators $\S_1$, $\S_2$, $\T$ 
span the complete 3-dimensional tangent space in a generic point $(x,y,z)$.
Precisely this requirement will be assumed to hold hereafter;
operators $L$ with this property will be called {\em generic}.

Let us fix the coefficients in the expansions of the following
commutators:
\begin{equation}\label{st-comm1}
        [\S_2, \T] = K(x,y,z)\S_1 +M(x,y,z)\S_2 + N(x,y,z)\T.
\end{equation}
\begin{equation}\label{st-comm2}
        [\S_1, \S_2] = P(x,y,z)\S_1 +Q(x,y,z)\S_2 + R(x,y,z)\T.
\end{equation}

First we try to represent the operator in a  partially factorized form: $L=(\S_1 + \a)(\S_2 +\b)  + \V + b(x,y,z)$  
with some indefinite $\a=\a(x,y,z)$, $\b=\b(x,y,z)$
and $\V=\T-\b\S_1 -\a\S_2$, $b=a-\a\b-\S_1(\b)$.

Then introducing $v=(\S_2 +\b)u$ we get the corresponding
first-order system:
\begin{equation}\label{Dini2}
Lu=0 \Longleftrightarrow
\left\lbrace  \begin{array}{l}
(\S_2 +\b)u = v, \\
(\V+b) u = - (\S_1 + \a)v.
\end{array}\right.
\end{equation}
Next we try to eliminate $u$
by cross-differentiating the left hand sides, which gives
\begin{equation}\label{th-L1}
[(\V+b), (\S_2 +\b) ]u = (\S_2 +\b)(\S_1 + \a)v +(\V+b)v.
\end{equation}
If one wants $u$ to disappear from this new equation one should find out 
when $[(\V+b), (\S_2 +\b)]u$ can be transformed into an expression
involving \emph{only} $v$, i.e.\ when this commutator 
is a linear combination of just two expressions
$(\S_2 +\b)$ and $(\V+b)$:
\begin{equation}\label{th-comm}
        [(\V+b), (\S_2 +\b)] = \mu(x,y,z)(\S_2 +\b) + \nu(x,y,z)(\V+b).
\end{equation}
This is possible to achieve choosing the free functions
$\a(x,y,z)$, $\b(x,y,z)$ appropriately. 
In fact, expanding the left and right hand sides in (\ref{th-comm})
in the local basis of the initial fixed operators $\S_1$, $\S_2$, $\T$
and the zeroth-order operator $1$
and collecting the coefficients of this expansion, one gets the following system
for the unknown functions $\a$, $\b$, $\mu$, $\nu$:
$$
 \left\lbrace  \begin{array}{l}
    K  +\b P - \S_2(\b) = \nu \b   ,     \\
    M  -\S_2(\a) +\b Q  =   \nu \a - \mu ,\\
    N + \b R = - \nu       ,  \\
    \b\S_1(\b)-\T(\b)+\S_2(a) -\b\S_2(\a) -\S_2(\S_1(\b)) = 
          - \nu (a-\a\b-\S_1(\b))-\mu\b.
 \end{array}\right.
$$
After elimination of $\nu$ from its first and third equations we get 
a first-order non-linear partial differential equation for $\b$:
\begin{equation}\label{eq-b}
 \S_2(\b) = \b^2R+(N+P)\b +K.
\end{equation}
This Riccati-like equation may be transformed into a second-order linear PDE
via the standard substitution $\b = \S_2(\gamma)/\gamma$.
Taking any non-zero solution $\b$ of this equation and substituting 
$\mu = \nu \a +\S_2(\a) -\b Q -M $ (taken from the second equation of the system) 
into the fourth equation of the system we obtain
a first-order linear partial differential equation for $\a$
with the first-order term $\b\S_2(\a)$. Any solution of this equation
will give the necessary value of $\a$. Now we can substitute
$[(\V+b), (\S_2 +\b)]u = \mu(\S_2 +\b)u + \nu(\V+b)u=
\mu v -\nu (\S_1 + \a)v$ into the left hand side of (\ref{th-L1})
obtaining the transformed equation $L_{(1)}v=0$.

If we would start the same procedure using the second partial factorization in (\ref{th-1}) we would find the other transformed
equation  $L_{(-1)}w=0$.
 $\Box$

As a rule neither of the obtained new operators 
$L_{(1)}$, $L_{(-1)}$ factors into a product of first-order operators
as was the case for the operator $L = (\hat D_x\hat D_y + x \hat D_x\hat D_z - \hat D_z)$ in the previous section.
Then one can repeat the described process due to the fact that
the principal symbol of the transformed equations is still $\S_1\S_2$.
Thus we have in the case treated in this section an infinite chain of {\em Dini transformations}
$$  
 \ldots {\leftarrow} \quad L_{(-2)} \quad {\leftarrow}\quad 
    L_{(-1)}\quad {\leftarrow}  \quad   L  \quad {\rightarrow}\quad 
     L_{(1)} \quad {\rightarrow} \quad  L_{(2)}\quad  {\rightarrow}
     \ldots
$$  
If  some of the $L_{(i)}$ is factorizable we can obtain its complete
solution (under the assumption that one can solve the corresponding
first-order equations explicitly) and solving the system (\ref{Dini2})
w.r.t.\ $u$ step through this chain back (this again requires solution of linear first-order equations) finally obtaining the complete solution of the original equation
$Lu=0$.

\section{Open problems}
\label{concl}

In the previous three sections we tacitly assumed that the problem of solution of \emph{first-order} linear equations
\begin{equation}\label{1lin1}
  \left(\sum_{i}b_{i}(\vec x)\hat D_{x_{i}} +b_{0}(\vec x)\right) u =0
\end{equation}
or
\begin{equation}\label{1lin2}
  \left(\sum_{i}b_{i}(\vec x)\hat D_{x_{i}}  +b_{0}(\vec x)\right)u = \phi(\vec x)
\end{equation}
can be solved at least for polynomial $b_i(\vec x)$. In fact it is well known that even in the case $dim=2$ 
the problem of complete solution of $(b_1(x,y)\hat D_x + b_2(x,y)\hat D_y)u=0$
is equivalent to finding a nontrivial conservation law for the corresponding nonlinear autonomous ODE system or a non-autonomous first-order ODE:
\begin{equation}\label{12sys}
\left\lbrace  \begin{array}{l} 
  \displaystyle \frac{dx}{dt} = b_1(x,y), \\[2ex]
  \displaystyle \frac{dy}{dt} = b_2(x,y),
\end{array}\right.    \ \ \ \ \Longleftrightarrow \ \ \  \frac{dy}{dx} =\frac{b_2(x,y)}{b_1(x,y)}
\end{equation}
(or, equivalently, finding their general solutions). For polynomial $b_i(x,y)$ 
this is one of the famous fields of research: study of polynomial vector fields in the plane. Recently an essential advance was made in \cite{chen-ma,ere98,feng-gao}; 
one may hope that a complete algorithm may be found. 
Still the problem of finding complete solutions of 
(\ref{1lin1})  in a suitable ``constructive´´ differential field
algorithmically is a challenging problem, as well as the problem of finding solutions for
the equation (\ref{eq-b}).

Another challenging problem is to establish a connection between the general theoretic definition given in \cite{ts98} and the exposed above practical methods based of Laplace and Dini transformations.
The known cases suggest the following conjectures presumably valid
for operators of arbitrary order and any number of independent variables:

\begin{itemize}
        \item \textbf{Conjecture 1}. If a LPDO is factorizable in the generalized sense of \cite{ts98}, then its
principal symbol is factorizable as a commutative polynomial in formal variables $\hat D_{x_i}$.

\medskip

\item
\textbf{Conjecture 2}.
If a LPDO of order $n$ is solvable (i.e.\ the corresponding linear homogeneous equation has an explicit closed-form solution) then its principal symbol splits into 
product of $n$ linear factors.
\end{itemize}

One may also suggest to define the principal symbol of a LPDO using different weights for different $\hat D_{x_i}$; this would imply for example
generalized irreduciblity of parabolic operators similar to $\hat D_x^2-\hat D_y$
and potentially provide a powerful criterion of (un)solvability.

One should point out that the \emph{methods} for solution of LPDEs given in the previous sections can not be called completely \emph{algorithmic}: even for the classical case of Laplace transformations and the simplest possible characteristic operators $\hat X_1=\hat D_x$, $\hat X_2=\hat D_y$ we do not have any bound on the  number of steps in the chain (\ref{ch}). Example~1 given in Section~\ref{lap} suggests that 
such bounds or other hypothetic stopping criteria would depend on rather fine arithmetic properties of the coefficients of LPDOs.

A more general theoretic treatment suitable for arbitrary (even under-  or over-determined systems, cf.\ \cite{L-Sch-Ts,minwu}) based on the language of  abelian categories will be exposed in a later publication.

A link to the theory of Darboux-integrable nonlinear PDEs established in
\cite{anderson,sokolov,S-Zh} in our opinion can be extended to other 
types on nonlinear PDEs. In this connection a generalization of Laplace invariants for higher-dimensional and higher-order cases started in \cite{Athorne}, \cite{YA}, \cite{ts05}, \cite{SW} would be of extreme importance.

\section*{Acknowledgments}

The author enjoys the occasion to thank the WWCA-2006 organizers
for their efforts which guaranteed the success of the Workshop as
well as for partial financial support which made presentation of the results given above possible. Certainly the 
 honor of being co-author of a joint paper with S.A.~Abramov and our numerous  discussions about algorithmic solution of differential equations were (and will always be) highly inspiring!

\end{document}